\documentclass[sigconf, nonacm]{acmart}

\usepackage{multirow} 
\usepackage{array, ragged2e}    
\usepackage{booktabs} 
\usepackage{mdframed}
\usepackage{fontawesome5}
\usepackage{enumitem}
\setlist[itemize]{leftmargin=12pt}
\usepackage[skip=4pt]{caption}

\begin{document}

\title{Toward Agentic Software Engineering Beyond Code:\\ Framing Vision, Values, and Vocabulary}

\author{Rashina Hoda}
\orcid{0000-0001-5147-8096}
\affiliation{%
  \institution{Monash University}
  \city{Melbourne}
  \country{Australia}}
  \email{rashina.hoda@monash.edu}

\renewcommand{\shorttitle}{Toward Agentic SE Beyond Code}
\renewcommand{\shortauthors}{Hoda}

\begin{abstract}
Agentic AI is poised to usher in a seismic paradigm shift in Software Engineering (SE). As technologists rush head-along to make \textit{agentic AI} a reality, SE researchers are driven to establish \textit{agentic SE} as a research area. While early visions of agentic SE are primarily focused on code-related activities, early empirical evidence calls for a consideration of a wider range of socio-technical activities and concerns to make it work in practice. This paper contributes to the emerging visions by: (a) recommending an expansion of its scope beyond code, toward a `whole of process' \textit{vision}, grounding it in SE foundations and evolution and emerging agentic SE frameworks, (b) proposing a preliminary set of \textit{values and principles} to guide community efforts, and (c) sharing guidance on designing and using well-defined \textit{vocabulary} for agentic SE. It is hoped that these ideas will encourage collaborations and steer the SE community toward laying strong foundations of agentic SE so it is not limited to enabling coding acceleration but becomes the next process-level paradigm shift.

\end{abstract}


\maketitle

\section{Introduction}

The Software Engineering (SE) industry, and consequently the SE research discipline, is at the verge of an incoming seismic paradigm shift proposed by the highly anticipated advent of \textbf{agentic AI}. Latest releases of agentic systems (e.g., SWE-agent, Jules, Codex, Devin) are fast making agentic AI a reality. Following suit, visions capturing various components of \textbf{agentic SE} are being proposed, e.g., \textit{agentic AI software engineers} \cite{roychoudhury2025agentic}, \textit{unified software engineering agent} (USEagent) \cite{applis2025unified}, and \textit{structured agentic software engineering} (SASE) \cite{hassan2025agentic}. Li et al. have made \textit{AIDev dataset} comprising of over 450K agentic pull requests created by leading autonomous coding agents publicly available \cite{li2025rise}, positioning agentic SE as an ``unfolding reality''. \textbf{These visions, while necessary and welcome, primarily focus on and around one SE activity -- coding}, e.g., code generation, review, debugging, repair, configuration etc. Whereas other parts of the SE lifecycle, e.g., requirements engineering, design, operations, and practical aspects such as adoption, teamwork, management, and workflow integration also need attention.

\begin{mdframed}[innerleftmargin=2pt, innerrightmargin=2pt, innertopmargin=10pt, innerbottommargin=10pt, skipabove=20pt, skipbelow=30pt, linewidth=0.2pt]\footnotesize{Accepted to the AGENT workshop at the International Conference on Software Engineering (ICSE) 2026}
\end{mdframed}

Early empirical studies show AI's role as a ``personal accelerator'' for coding, writing, and documentation tasks but highlight its limitations in addressing teamwork, coordination, accountability, and culture \cite{xiao2025ai}. Similarly, organizational adoption challenges and human-AI collaboration have been highlighted as areas deserving research focus \cite{akbar2025agentic}. In a review of \textit{AI agentic programming}, Wang et al. highlight the increasing human and socio-technical concerns requiring interdisciplinary solutions \cite{wang2025ai}. Abrah{\~a}o et al. pose several socio-technical questions for the age of AI (e.g., role of human software engineers, how human-AI collaboration might work, how ``citizen engineers'' might interact with AI agents) \cite{abrahao2025software}. 

For agentic SE to become the next process-level paradigm shift (and not just a coding accelerator), we need to \textbf{expand the emerging vision of agentic SE beyond code} to examine a full range of socio-technical activities, artefacts, roles, and concerns across the entire SE lifecyle, with traditional ones renewed for agents and multi-agent systems as well as new ones required to make human-agent ecosystems work. This paper contributes to the emerging agentic SE vision by: (a) extending its scope beyond coding, toward a \textit{\textbf{`whole of process'}} \textbf{vision}, grounding it in a discussion of SE foundations and evolution (in Section \ref{models} and \ref{towardagenticSE}), (b) proposing a preliminary set of \textbf{values and principles} to guide our collective design of the emerging vision (in Section \ref{craft_principles}), and (c) highlighting the importance of, and sharing guidance on, building well-defined \textbf{vocabulary} (terminology) for the emerging agentic SE discipline (in Section \ref{terminology}). Together, it is hoped that these ideas will steer our emerging community vision of agentic SE toward laying the core foundations of agentic SE process models, more fleshed out values and principles, and well-defined vocabulary to support our conversations, in verbal and written forms. As with other recent visions, these are not meant to be exhaustive or final. Rather, they are \textbf{intended to encourage community feedback and discussions} about adopting a systematic and principled approach to thinking about, researching, and enabling agentic SE (i.e., agentic AI for SE, SE for agentic AI). 

\section{SE Foundations and Evolution}\label{models}
Before we can \textit{redefine} a field, it makes sense to revisit its foundational definition and evolution. SE is defined as ``\textit{the application of a systematic, disciplined, quantifiable approach to the development, operation, and maintenance of software; that is, the application of engineering to software.}'' (ISO/IEC/IEEE 24765:2017 Systems and Software Engineering — Vocabulary/Guide to SWEBOK 4.0a.

Ergo, \textbf{SE is more than coding.} Starting out as a ``cottage industry'' in the 1950s, SE became an established engineering discipline with many \textbf{SE process models} (AKA software development lifecycle (SDLC), frameworks, methods) proposed, adopted, established, and replaced over the decades \cite{hoda2018rise}. All of them featured a `\textbf{whole of process}' approach. Royce introduced structure and formality to SE in 1970 through a two-iteration approach called \textbf{\textit{Waterfall}}. It was widely adopted as a specification-driven, sequential approach in practice, featuring core SE \textit{activities} as phases (requirements, analysis, design, coding, testing, and operations), \textit{artefacts} (documents from each phase), and formalized \textit{roles} (requirements engineers, designers, developers, testers etc.). Over time, \textit{tools} and \textit{techniques} were introduced and refined to support the roles to implement the activities and produce the artefacts. Later, the \textbf{\textit{Sprial model}} and \textbf{\textit{V-model}} espoused a risk-driven and verification-validation-driven approach to SE, respectively. While, the \textit{\textbf{Rational Unified Process (RUP)}} highlighted software project management concerns such as timely delivery and budget \cite{kruchten2004rational}.

\textbf{\textit{Agile software development}} marked a paradigm shift in SE with its people-centeredness, collaborative approach, focus on working software, and ability to pivot in response to changes. Agile methods, \textit{eXtreme Programming} and \textit{Scrum}, came to dominate the SE industry \cite{hoda2018rise}. Agile SE researchers studied its newly introduced roles, practices, artefacts, its impact on related (sub)disciplines (e.g., requirements engineering, project management) and integration with seemingly contradictory approaches (e.g., CMMI, safety-critical systems) \cite{hoda2018rise}. Lean principles manifested through \textbf{\textit{Kanban}}. Extending the coverage of the SDLC, \textbf{\textit{DevOps}} connected development and operations. Agile became an industry default/standard \cite{stateOfAgile}.

The use of intelligent techniques to support various activities of the SE lifecycle has been of interest over the years \cite{perkusich2020intelligent}. In 2019, a conceptual framework of \textbf{\textit{AI/ML powered agile}} was proposed \cite{dam2019towards}. Building on that, a vision of \textit{\textbf{Augmented Agile}}, an AI-powered agile project management approach, combining the `human heart' (e.g., humans values and concerns) with `AI brains', imagined as an \textit{agile co-pilot} team member was proposed \cite{hoda2023augmented}. These techniques and conceptual ideas foreshadowed \textbf{\textit{GenAI/LLM powered SE practices}}, such as LLM-based multi-agent systems for SE \cite{he2025llm}.

A review of 395 papers on the use of LLM-based solutions for SE found that \textit{software development} (i.e., coding and related) was the most researched activity while human-intensive activities (e.g., RE, design, management) were the least explored \cite{hou2024large}. Recently, \textbf{\textit{AgileGen}}, focused on maintaining consistency between the generated software and end-user requirements was proposed \cite{zhang2025empowering}. Others have called for GenAI to be human-centered \cite{russo2024generative}.

\vspace{25mm}

\begin{figure}[]
    \centering
    \includegraphics[width=0.96\linewidth]{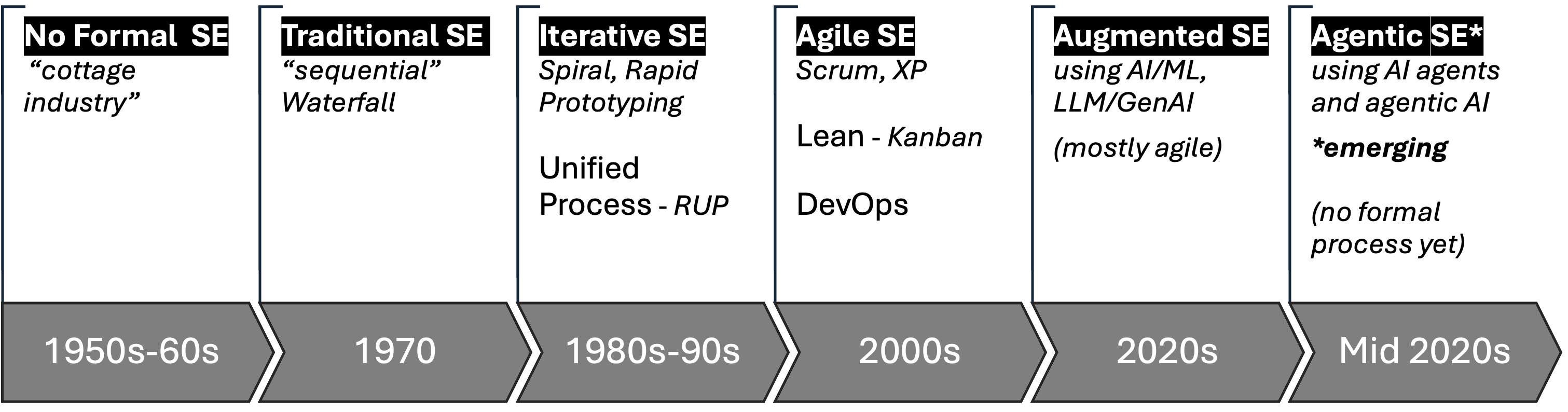}
    \caption{Software Engineering Evolution: A Brief Timeline}
    \label{fig:SEevolution}
    \Description{Software Engineering Evolution: A Brief Timeline}
\end{figure}

\begin{mdframed}[innerleftmargin=2pt, innerrightmargin=2pt, innertopmargin=2pt, innerbottommargin=2pt, skipabove=1pt, skipbelow=1pt, linewidth=1pt, backgroundcolor=black!5] \footnotesize{For agentic SE to become the next process-level paradigm shift in SE, we need to \textbf{expand our vision of agentic SE beyond code} to examine a full range of \textbf{areas} and \textbf{activities} (e.g., ethical alignment, RE, design, development, operations), associated \textbf{actors} (e.g., teams of human and AI software engineers, end-users), \textbf{artefacts} (e.g., code, data, evidence and analyses samples, experimental infrastructure), and other \textbf{socio-technical concerns} -- traditional SE ones renewed for agents and multi-agent systems, and new ones needed to make human-agent ecosystems work. The establishment of SE can be credited to the formalization of various \textbf{process models}, the development of supporting \textbf{tools}, \textbf{techniques}, and the formalization of \textbf{knowledge areas} and \textbf{vocabulary}. We need the same to establish \textbf{agentic SE}.}
\end{mdframed}

\section{Toward Agentic SE (emerging)}\label{towardagenticSE}

Google’s Jules, OpenAI’s Codex, Cognition’s Devin, SWE-agent, AutoCodeRover, and Anthropic’s Claude Code are making agentic AI a reality. Unsurprisingly, visions of new agentic SE frameworks and approaches are being proposed. In an opinion piece, Roychoudhury et al. propose a new \textbf{role} of \textbf{\textit{agentic AI Software Engineer}}, an  autonomous LLM-based coding agent ``\textit{that combines coding, testing, debugging, etc. into a coherent, explainable workflow}'' \cite{roychoudhury2025agentic}. They predict a shift in interest from \textit{programming in the large} to \textit{programming with trust} and highlight the importance of operationalizing socio-technical integration of AI agent trust, predicting a future where human developers can rely on autonomous agents to become trusted programming collaborators. The idea of a \textbf{\textit{unified software engineering agent (USEagent)}} with capabilities to handle multiple programming tasks (e.g., CodeRetrieval, ExecuteTests, EditCode, ReviewPatch) is described in \cite{applis2025unified}. It promises to be ``\textit{a first draft of a future AI Software Engineer which can be a team member in future software development teams involving both AI and humans.}'' 

Hassan et al. propose \textbf{\textit{structured agentic software engineering}} (SASE) as a conceptual framework for ``two symbiotic modalities'', namely, \textit{SE for Humans} and \textit{SE for Agents} \citep{hassan2025agentic}. They propose a hierarchical framework with increasing levels of autonomy: from manual coding (level 0) to goal-agentic (dubbed Agentic SE/SE 3.0) and all the way to general domain autonomy (level 5, SE 5.0). They envision humans and agents engaging in agent-based software programming \textit{activities} supported by dedicated agent command environment (ACE) and agent execution environment (AEE). They share a vision of agents driving software programming, supported by evolving artefacts (e.g., briefingpacks), roles (e.g., humans as coaches and mentors for agents) and tools (e.g., ACE and AEE replacing traditional IDEs). Recently, Li et al. made \textbf{\textit{AIDev}}, a large-scale dataset of over 450K PRs by leading autonomous agents, publicly available for SE and AI researchers to longitudinally study a range of research questions, e.g., to do with agent behaviour and dynamics, human-AI collaboration, productivity, effort, quality etc. \cite{li2025rise}.  \textbf{These visions and efforts, while necessary and welcome, are primarily focused around one SE activity -- coding}. 

Besides the technical challenges of implementing agentic AI for programming-related SE tasks, a number of \textbf{\textit{socio-technical concerns across the SDLC}} arise. Early empirical evidence highlights the need to address agentic AI applicability across the SE process. Xiao et al.'s longitudinal study showed AI's role as `a personal accelerator' but it was not seen to fix \textit{teamwork issues} and its impact on \textit{coordination}, \textit{accountability}, and \textit{culture} remain to be seen \cite{xiao2025ai}. Akbar et al. report that current studies ``often focus narrowly on AI-powered code generation or bug detection without addressing'': \textit{cross-phase applicability}, \textit{organizational adoption challenges}, and \textit{human-AI collaboration patterns} \cite{akbar2025agentic}. Similarly, Wang et al. review \textbf{\textit{AI agentic programming}} and highlight that increasing human and socio-technical concerns with AI agentic programming such as \textit{context handling}, \textit{human-AI collaboration}, \textit{verification of agent behaviour}, will require ``\textit{interdisciplinary solutions}'' \cite{wang2025ai}. Similar socio-technical issues (e.g., \textit{human-AI collaboration}, \textit{next generation educatio}n, \textit{job security}, \textit{ethics}, \textit{diversity}) were highlighted by Abrah{\~a}o et al. \cite{abrahao2025software}. How \textbf{nascent} agentic SE is can be gauged from the fact that most of these proposals are under review (Jan 2026).

\subsection*{A `Whole of Process' Vision}
Agile was a \textit{process} revolution, defining new SE roles, practices, and artefacts \cite{hoda2018rise}. Agentic SE is a \textit{technology} revolution. Currently, there is no guidance on how AI teammates can/should be incorporated across the SDLC. Building on the foundational aspects of previous SE models (Section \ref{models}) and AI capabilities of the proposed frameworks (Section \ref{towardagenticSE}), this paper presents a \textit{preliminary `whole of process'} \textbf{vision for agentic SE }(Figure~\ref{fig:SEevolution}). It is not meant to be a definitive agentic SE process model, rather a `first look' that will need to be improved and expanded on. The vision extends across high-level core \textbf{areas} of the SE lifecycle, approached iteratively, by human and agent (single agents or multi-agent system) \textbf{actors}, collaborating in \textbf{activities} at varying levels of AI agency (depicted by the slider under \textit{human control} \cite{spiegler2025images}). Given AI's increasing impact on society, it introduces \textit{ethical alignment} as first and central.

\begin{itemize}
    \item \textbf{Agentic SE Ethical Alignment}: covers activities, actors, and artefacts (AAA) to do with ethics alignment. With increasing AI capabilities and agency and their potential for wide-scale societal impact \cite{spiegler2025images}, alignment with safe AI laws and regulations at the \textit{micro} (e.g., project, company) and \textit{macro} (e.g., domain, region, country, international) levels is a must.
    \item \textbf{Agentic Requirements Engineering}: covers AAA to do with software requirements (e.g., customer, user, subject matter expert, quality, and cascading ethical requirements.) 
    \item \textbf{Agentic Design}: covers AAA to do with software design (e.g., architecture, UI, UX, design for human benefit and equity). 
    \item \textbf{Agentic Development}: covers AAA to do with software code design, production, testing, reviewing, repairing, etc. We have a head-start on envisioning the transformation in this area with emerging visions and an agentic AI dataset \cite{roychoudhury2025agentic, roychoudhury2025thoughts, hassan2025agentic, li2025rise}.
    \item \textbf{Agentic Operations}: covers AAA to do with deployment, maintenance, and operations (e.g., Ops/DevOps/DevSecOps).
\end{itemize}

These high-level area descriptions will need to be fleshed out in collaboration with area experts, including relevant RQs, example scenarios, and specific challenges. Transforming traditional SE areas and introducing new ones (e.g., ethical alignment) will ensure agentic SE leverages the full capabilities of agents to support humans and improve outcomes across a full range of SE activities.

\begin{figure}[t]
    \centering
    \includegraphics[width=1\linewidth]{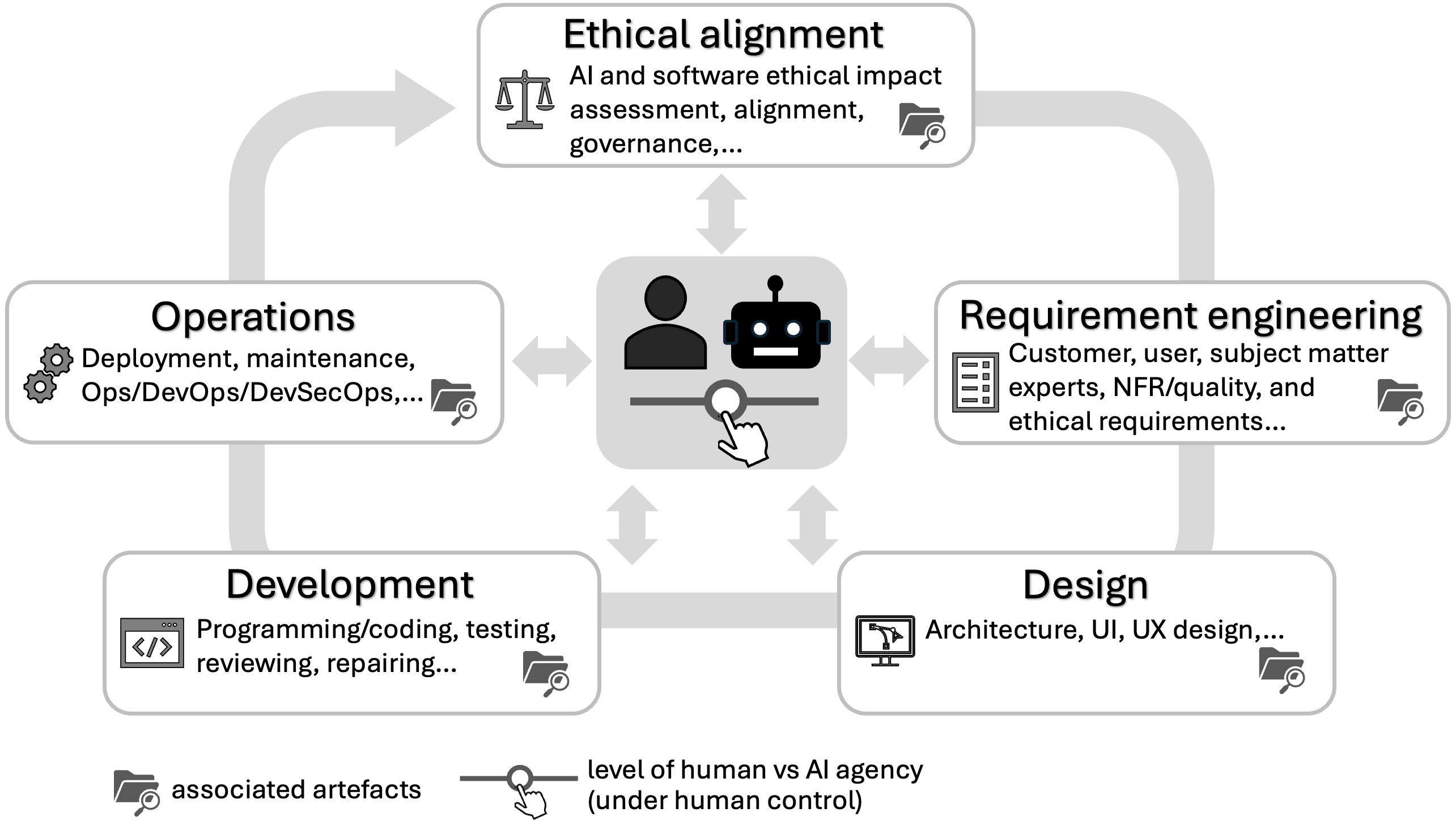}
    \caption{A `whole of process' vision for Agentic SE, where human and agent actors iterate across areas, activities, artefacts with varying levels of AI agency (under human control)}
    \label{fig:SEevolution}
    \Description{A `whole of process' vision for Agentic SE}
\end{figure}

\section{Toward Agentic SE Values and Principles} \label{craft_principles}
To guide this paradigm shift, a set of preliminary \textit{comprehensive}, \textit{responsible}, \textit{adaptive}, \textit{foundational}, and \textit{translational} \textbf{(CRAFT) values and principles for agentic SE} is presented to spark community conversation and collaboration. Most apply to both research (\faFlask) and practice (\faCogs), while some are primarily research focused.

\subsubsection*{\textbf{Value 1. Comprehensive:}} Agentic SE needs to take a \textit{`whole of process approach'} and include the consideration of \textit{human and socio-technical aspects} in addition to core technological innovation.

\begin{table*}[h!]
    \centering
    \scriptsize
    \caption{CRAFT Values and Principles for Agentic Software Engineering (Preliminary)}
    \label{tab:craft}
    \begin{tabular}{>{\raggedright}p{1.3cm} >{\raggedright}p{2.2cm} p{13cm}}
    \toprule
    \textbf{Value} & \textbf{Principle} & \textbf{For Agentic SE}\\
    \midrule
    \multirow[t]{2}{*}{\textbf{C}omprehensive}
         & \textit{Whole of Process Approach} & Agentic SE needs to take a ‘whole of process’ approach to examining traditional SE areas and associated activities, actors, artefacts (AAA) and consider new ones to enable emerging human-agent ecosystems.\\
         & \textit{Human, Agentic, and Socio-technical Aspects} & Agentic SE needs to include a focus on human, agentic, and socio-technical aspects of SE including \textit{roles} played by human and agent actors, \textit{human and agentic aspects} (e.g., human values, empathy, motivation, emotional intelligence, personality) and socio-technical aspects (e.g., mutual trust, adoption barriers, teamwork, collaboration, communication, coordination, project management).\\

    \multirow[t]{2}{*}{\textbf{R}esponsible}
         & \textit{Ethics-by-Design} & Agentic SE needs to apply an \textit{ethics-by-design} approach to its design and application, questioning the \textit{why}, and not being driven by the \textit{why not}.\\
         & \textit{Sustainability} & Agentic SE needs to ensure sustainability of the core technologies, ecosystems, and the planet.\\

    \multirow[t]{2}{*}{\textbf{A}daptive}
         & \textit{Industrial Relevance} & Agentic SE needs to adapt to new AI models and paradigms and work closely with the AI community and co-design with the industry.\\
         & \textit{Socio-technical Relevance} & Agentic SE needs to predict and report socio-technical impacts of changes in AI models and paradigms.\\

    \multirow[t]{2}{*}{\textbf{F}oundational}
         & \textit{Foundational Knowledge} & Agentic SE needs to develop foundational knowledge about AAA across all areas, not only coding/programming.\\
         & \textit{Foundational Solutions} & Agentic SE needs to develop foundational solutions to support AAA across all areas, not only coding/programming.\\

    \multirow[t]{2}{*}{\textbf{T}ranslational}
         & \textit{Awareness and Education} & Agentic SE needs to build awareness through education, certifications, funding bodies, media, and society; and maintain incident logs.\\
         & \textit{Actionable Guidelines} & Agentic SE needs to study and enable the adoption of research knowledge and solutions into practice through usable guidelines and tools.\\
         
    \bottomrule
    \end{tabular}
\end{table*}

\begin{itemize}
    \item \textit{\textbf{Principle of Whole of Process \faFlask \faCogs}} -- \textit{Agentic SE needs to take a `whole of process' approach} -- A `whole of process' approach, a hallmark of previous SE models (section \ref{models}). Perhaps being one of the reasons why \textit{Scrum} (covering the full SE lifecycle and project management aspects) considerably outperformed \textit{extreme programming} \cite{hoda2018rise}. Relevant parts of the SE lifecycle, not only its code-related activities, need to be considered as well as new required activities, in designing agentic SE. This includes assessing the goals and constrains related to ideation, RE activities (e.g., to establish user needs and preferences), design, architecture, project management (e.g., costs, times, resources), functional and user testing, delivery, maintenance, operations/evolution, as well as for new practices, trade-offs, and  optimizations. 
    
    \item \textit{\textbf{Principle of Human, Agentic, and Socio-technical Aspects \faFlask \faCogs}} -- \textit{Agentic SE needs to include the human, agentic, and socio-technical aspects of SE} -- Like SE, agentic SE is socio-technical, where ``\textit{the social and technical aspects are inherently interwoven}'' \cite{hoda2024qualitative}. Technical innovation is necessary but not sufficient for advancing agentic SE, e.g., trust and utility concerns outrank performance in real-world adoption \cite{li2025rise}. Human, agentic, and socio-technical aspects must be considered, e.g., the \textit{roles} played by human and agent actors, \textit{human and agentic aspects} (e.g., human values, empathy, emotional intelligence), and \textit{socio-technical aspects} (e.g., mutual trust, adoption barriers, human experience, coordination, communication, collaboration, conflict management, teamwork). Research on these will help the community answer fundamental questions such as: \textit{What should be automated, to what extent (human oversight levels), and how? What shouldn't be automated and why? How should agentic systems be best integrated into teams and workflows?}
\end{itemize}

\subsubsection*{\textbf{Value 2. Responsible:}} Agentic SE needs to ensure \textit{ethics-by-design} and \textit{sustainability}. With agentic AI injecting more agency into AI, we need positive societal impact and human control of AI \cite{spiegler2025images}. 
\begin{itemize}
    \item \textbf{\textit{Principle of Ethics-by-Design \faFlask \faCogs}} -- \textit{Agentic SE needs to apply an ethics-by-design approach to its design and application} -- There is unprecedented growth in AI-based companies (e.g., OpenAI, NVIDIA, Microsoft, Cursor) with claims of 100x productivity\footnote{\tiny{\url{https://www.businessinsider.com/surge-ceo-ai-100x-engineers-2025-7}}}. \textit{But to what end and at what cost?} It is critical to take an \textit{ethics first} approach to agentic SE, where ethical aspects (e.g., accountability, reliability, safety, transparency, explainability, privacy, security, human/societal/environmental wellbeing \cite{lu2022software}) are prioritized. Applying \textit{ethical research}, we should question the \textit{why} and not be driven by the \textit{why not} \cite{storey2025guiding} (e.g., \textit{why} does \textit{X} need to be automated vs. why not automate \textit{X}).
    
    \item \textbf{\textit{Principle of Sustainability \faFlask \faCogs}} -- \textit{Agentic SE needs to ensure sustainability} -- In addition to ensuring ultimate human control and positive social impact \cite{spiegler2025images}, the planetary cost of AI technologies cannot be ignored \cite{google2025energy}. As Cruz et al. highlight in their article on \textit{Green AI}, ``\textit{our [SE] community needs to adopt a long-term plan enabling a conscious transformation that aligns with environmental sustainability values}'' \cite{cruz2025innovating}, which necessitates handling trade-offs \cite{lago2015framing}.
\end{itemize}

\subsubsection*{\textbf{Value 3. Adaptive:}} Agentic SE needs to adapt to maintain \textit{industrial} and \textit{socio-technical} relevance. 
\begin{itemize}
    \item \textbf{\textit{Principle of Industrial Relevance \faFlask \faCogs}} -- \textit{Agentic SE researchers and practitioners need to adapt to new AI models and paradigms} -- \textbf{Agentic SE fundamentally ties SE to AI.} Agentic SE and AI communities should work in tandem, adapting and contributing to the latest AI models and paradigms. Empirical studies of problem and solution spaces will be useful in maintaining industrial relevance. Co-designing and evaluations with industry will help.

    \item \textbf{\textit{Principle of Socio-technical Relevance \faFlask \faCogs}} -- \textit{Agentic SE needs to predict and report socio-technical impacts of changes in AI models and paradigms} -- Maintaining \textit{relevance} extends to the associated human (e.g., evolving levels of AI agency and human control \cite{spiegler2025images}) and socio-technical aspects (e.g., awareness, ethics, adoption). This implies anticipating and keeping up with the impacts of changing AI technology and envisioning, studying, and reporting them promptly to inform research and practice.
    
\end{itemize}

\subsubsection*{\textbf{Value 4. Foundational:}} Agentic SE needs to develop foundational \textit{knowledge} and \textit{solutions} beyond coding.
\begin{itemize}
    \item \textbf{\textit{Principle of Foundational Knowledge \faFlask}} -- \textit{Agentic SE needs to develop foundational \textit{knowledge} beyond coding} -- As with agile methods \cite{hoda2018rise}, establishing foundational knowledge through empirical studies of new roles, practices, artefacts, and socio-technical aspects  will benefit agentic SE. Building original taxonomies and theories \cite{hoda2024qualitative} and testing existing ones will contribute to theoretical foundations, e.g., \textit{agents in SE taxonomy} \cite{wang2025agents}. Qualitative \cite{seaman2025qualitative}, quantitative, and mixed methods studies \cite{storey2025guiding}, and intra/inter-disciplinary work will help build a body of knowledge. 
    
    \item \textbf{\textit{Principle of Foundational Solutions \faFlask \faCogs}} -- \textit{Agentic SE needs to develop foundational \textit{solutions} beyond coding} -- Proposed visions of programming frameworks are emerging \cite{hassan2025agentic, roychoudhury2025agentic}. As with GenAI/LLM based solutions (tools, techniques, frameworks, platforms) for SE \cite{hou2024large}, foundational solutions for agentic SE are expected to be proposed. It will be critical to extend such solutions to support other agentic SE areas beyond coding/programming.
\end{itemize}

\subsubsection*{\textbf{Value 5. Translational:}} Agentic SE needs to enable research translation into practice through building \textit{awareness}, \textit{education}, and \textit{actionable guidelines}.
\begin{itemize}
    \item \textbf{\textit{Principle of Awareness and Education \faFlask \faCogs}} -- \textit{Agentic SE needs to build awareness} -- While awareness of agentic AI is on the rise thanks to agentic AI systems, products, and news (including ethical incidents), awareness of agentic SE (as a discipline) will need to be built through education and training (in classrooms and workplaces) as well as with accreditation, certification, funding bodies, media, and society. Maintaining an agentic SE incident log will also help. Previous SE models have shown the importance of such efforts to the long-term future of those process models and their associated communities.
    
    \item \textbf{\textit{Principle of Usable Guidelines and Tools \faFlask}} -- \textit{Agentic SE needs to study and enable the adoption of research knowledge and solutions into practice} -- Building knowledge and solutions is necessary but not sufficient to progress agentic SE. To achieve real-world impact, they need to be translated into actionable guidelines and tools for practice. Mirroring and leveraging AI's multi-modal generation, they can have more industry-friendly interfaces and modes of delivery (e.g., posters, executive summaries, graphical abstracts, short videos).
\end{itemize}

\section{Toward Agentic SE Vocabulary} \label{terminology}

Well-defined terminology is a precursor to formalizing \textit{taxonomy} (e.g. Sapkota et al. distinguish between \textit{AI agents} and \textit{agentic AI} and present a taxonomy of AI agent paradigms \cite{sapkota2025ai}) and ultimately, \textit{knowledge areas}. Although agentic SE is nascent, drifts and discrepancies in terminology usage are already apparent. Some are syntactic (e.g., agentic AI software engineer vs. AI software engineer vs. agentic software engineer). While others are reflective of deeper semantic and even philosophical issues. So we can lay solid foundations for agentic SE, here are some \textbf{considerations for designing and using terminology for agentic SE}: 
\begin{itemize}
    \item \textbf{Relevance} -- \textit{fit for purpose and domain context} -- to what extent is the term relevant to the domain? E.g., traditional SE vocabulary can be revisited for relevance in the agentic SE context.
    
    \item \textbf{Coverage} -- \textit{conceptual completeness and accuracy} -- to what extent does the term cover (nearly) \textit{all} the goals and responsibilities of the agentic SE process area, actor, activity, or artefact it aims to support? For example, a software engineer is more than a coder, so an \textit{autonomous coding agent} \cite{li2025rise} needs to be able to do more than coding to be termed an \textit{AI software engineer}.
    
    \item \textbf{Acceptance} -- \textit{recognition and adoption by the community} -- to what extent will humans (especially in the role) agree to an AI agent being given their professional title? For example, people might judge human and robot paintings as \textit{art} but are less willing to see robots as \textit{artists} \cite{mikalonyte2022can}.
    \item \textbf{Consistency} -- \textit{uniformity of use, stability, adaptability} -- to what extent do different technology and research groups in the SE world use the terminology in the same way, both in form (syntax) and meaning (semantics)? how long do they remain stable? how do they adapt to changes over time?
    
    \item \textbf{Philosophical alignment} -- \textit{reflexivity leading to awareness of underlying assumptions, power dynamics, fairness} -- reflecting on who comes up with terminology and why? Who stands to gain, who stands to lose, and does everyone have a voice (e.g., global south, underfunded regions)? How can we ensure fairness?
\end{itemize} 

\section{A Deliberate and Desirable  Paradigm Shift} \label{conclusion}

Building on the foundational aspects of previous SE models (Section \ref{models}) and recent visions of agentic SE (e.g., \cite{hassan2025agentic, roychoudhury2025agentic}), this paper proposes a `\textbf{\textit{whole of process}}' \textbf{vision} that accounts for human and agent \textit{actors}, \textit{artefacts}, and \textit{activities} and other \textit{socio-technical concerns} across all \textit{areas} of agentic SE: traditional SE ones renewed for agents and multi-agent systems and new ones for human-agent ecosystems. It presents a \textit{preliminary} set of \textit{comprehensive}, \textit{responsible}, \textit{adaptive}, \textit{foundational}, and \textit{translational} \textbf{(CRAFT) values and principles for agentic SE}. It  offers guidance on designing and using agentic SE \textbf{vocabulary}, welcoming community collaboration to progress ideas. Inevitably, proposed ideas end-up being defined by practice (e.g., Waterfall's sequential fate). We may not know what agentic SE truly looks like until it is studied empirically `in the wild'. Longitudinal studies to find useful patterns in agentic repositories and forums, and in-depth empirical studies based on interviews, surveys, and experiments with agentic SE teams, using inter- and intra-disciplinary approaches, are needed.

\vspace{1mm}
\noindent \textbf{Acknowledgment} Thanks to John Grundy, Hashini Gunatilake, Abhik Roychoudhury, and the reviewers for their valuable feedback.

\makeatletter
\def\@ACM@printyear#1{}
\makeatother
\bibliographystyle{ACM-Reference-Format}
\bibliography{references}


\begin{thebibliography}{30}


\ifx \showCODEN    \undefined \def \showCODEN     #1{\unskip}     \fi
\ifx \showISBNx    \undefined \def \showISBNx     #1{\unskip}     \fi
\ifx \showISBNxiii \undefined \def \showISBNxiii  #1{\unskip}     \fi
\ifx \showISSN     \undefined \def \showISSN      #1{\unskip}     \fi
\ifx \showLCCN     \undefined \def \showLCCN      #1{\unskip}     \fi
\ifx \shownote     \undefined \def \shownote      #1{#1}          \fi
\ifx \showarticletitle \undefined \def \showarticletitle #1{#1}   \fi
\ifx \showURL      \undefined \def \showURL       {\relax}        \fi
\providecommand\bibfield[2]{#2}
\providecommand\bibinfo[2]{#2}
\providecommand\natexlab[1]{#1}
\providecommand\showeprint[2][]{arXiv:#2}

\bibitem[Abrah{\~a}o et~al\mbox{.}(2025)]%
        {abrahao2025software}
\bibfield{author}{\bibinfo{person}{Silvia Abrah{\~a}o}, \bibinfo{person}{John Grundy}, \bibinfo{person}{Mauro Pezz{\`e}}, \bibinfo{person}{Margaret-Anne Storey}, {and} \bibinfo{person}{Damian~A Tamburri}.} \bibinfo{year}{2025}\natexlab{}.
\newblock \showarticletitle{Software Engineering by and for Humans in an {AI} Era}.
\newblock \bibinfo{journal}{\emph{ACM Transactions on Software Engineering and Methodology}} \bibinfo{volume}{34}, \bibinfo{number}{5} (\bibinfo{year}{2025}), \bibinfo{pages}{1--46}.
\newblock


\bibitem[Akbar et~al\mbox{.}(2025)]%
        {akbar2025agentic}
\bibfield{author}{\bibinfo{person}{Muhammad~Azeem Akbar}, \bibinfo{person}{Arif~Ali Khan}, \bibinfo{person}{Muhammad Hamza}, \bibinfo{person}{Abdullah Ghaffar}, {and} \bibinfo{person}{Arash Hajikhani}.} \bibinfo{year}{2025}\natexlab{}.
\newblock \showarticletitle{Agentic {AI} in Software Engineering: Practitioner Perspectives Across the Software Development Life Cycle}.
\newblock \bibinfo{journal}{\emph{Available at SSRN: https://ssrn.com/abstract=5520159}} (\bibinfo{year}{2025}).
\newblock


\bibitem[Applis et~al\mbox{.}(2025)]%
        {applis2025unified}
\bibfield{author}{\bibinfo{person}{Leonhard Applis}, \bibinfo{person}{Yuntong Zhang}, \bibinfo{person}{Shanchao Liang}, \bibinfo{person}{Nan Jiang}, \bibinfo{person}{Lin Tan}, {and} \bibinfo{person}{Abhik Roychoudhury}.} \bibinfo{year}{2025}\natexlab{}.
\newblock \showarticletitle{Unified Software Engineering agent as {AI} Software Engineer}.
\newblock \bibinfo{journal}{\emph{arXiv preprint arXiv:2506.14683}} (\bibinfo{year}{2025}).
\newblock


\bibitem[Crownhart(2025)]%
        {google2025energy}
\bibfield{author}{\bibinfo{person}{C. Crownhart}.} \bibinfo{year}{2025}\natexlab{}.
\newblock \bibinfo{booktitle}{\emph{In a first, Google has released data on how much energy an {AI} prompt uses}}.
\newblock \bibinfo{type}{{T}echnical {R}eport}. \bibinfo{institution}{MIT}.
\newblock
\urldef\tempurl%
\url{https://www.technologyreview.com/2025/08/21/1122288/google-gemini-ai-energy/}
\showURL{%
\tempurl}


\bibitem[Cruz et~al\mbox{.}(2025)]%
        {cruz2025innovating}
\bibfield{author}{\bibinfo{person}{Lu{\'\i}s Cruz}, \bibinfo{person}{Xavier Franch}, {and} \bibinfo{person}{Silverio Mart{\'\i}nez-Fern{\'a}ndez}.} \bibinfo{year}{2025}\natexlab{}.
\newblock \showarticletitle{Innovating for Tomorrow: The Convergence of Software Engineering and Green {AI}}.
\newblock \bibinfo{journal}{\emph{ACM Transactions on Software Engineering and Methodology}} \bibinfo{volume}{34}, \bibinfo{number}{5} (\bibinfo{year}{2025}), \bibinfo{pages}{1--13}.
\newblock


\bibitem[Dam et~al\mbox{.}(2019)]%
        {dam2019towards}
\bibfield{author}{\bibinfo{person}{Hoa~Khanh Dam}, \bibinfo{person}{Truyen Tran}, \bibinfo{person}{John Grundy}, \bibinfo{person}{Aditya Ghose}, {and} \bibinfo{person}{Yasutaka Kamei}.} \bibinfo{year}{2019}\natexlab{}.
\newblock \showarticletitle{Towards effective {AI}-powered agile project management}. In \bibinfo{booktitle}{\emph{2019 IEEE/ACM 41st international conference on software engineering (ICSE-NIER)}}. IEEE, \bibinfo{pages}{41--44}.
\newblock


\bibitem[Hassan et~al\mbox{.}(2025)]%
        {hassan2025agentic}
\bibfield{author}{\bibinfo{person}{Ahmed~E Hassan}, \bibinfo{person}{Hao Li}, \bibinfo{person}{Dayi Lin}, \bibinfo{person}{Bram Adams}, \bibinfo{person}{Tse-Hsun Chen}, \bibinfo{person}{Yutaro Kashiwa}, {and} \bibinfo{person}{Dong Qiu}.} \bibinfo{year}{2025}\natexlab{}.
\newblock \showarticletitle{Agentic Software Engineering: Foundational Pillars and a Research Roadmap}.
\newblock \bibinfo{journal}{\emph{arXiv preprint arXiv:2509.06216}} (\bibinfo{year}{2025}).
\newblock


\bibitem[He et~al\mbox{.}(2025)]%
        {he2025llm}
\bibfield{author}{\bibinfo{person}{Junda He}, \bibinfo{person}{Christoph Treude}, {and} \bibinfo{person}{David Lo}.} \bibinfo{year}{2025}\natexlab{}.
\newblock \showarticletitle{LLM-Based Multi-Agent Systems for Software Engineering: Literature Review, Vision, and the Road Ahead}.
\newblock \bibinfo{journal}{\emph{ACM Trans. on Software Engineering and Methodology}} \bibinfo{volume}{34}, \bibinfo{number}{5} (\bibinfo{year}{2025}), \bibinfo{pages}{1--30}.
\newblock


\bibitem[Hoda(2024)]%
        {hoda2024qualitative}
\bibfield{author}{\bibinfo{person}{Rashina Hoda}.} \bibinfo{year}{2024}\natexlab{}.
\newblock \bibinfo{booktitle}{\emph{Qualitative research with socio-technical grounded theory}}.
\newblock \bibinfo{publisher}{Springer}.
\newblock
\urldef\tempurl%
\url{https://link.springer.com/book/10.1007/978-3-031-60533-8}
\showURL{%
\tempurl}


\bibitem[Hoda et~al\mbox{.}(2023)]%
        {hoda2023augmented}
\bibfield{author}{\bibinfo{person}{Rashina Hoda}, \bibinfo{person}{Hoa Dam}, \bibinfo{person}{Chakkrit Tantithamthavorn}, \bibinfo{person}{Patanamon Thongtanunam}, {and} \bibinfo{person}{Margaret-Anne Storey}.} \bibinfo{year}{2023}\natexlab{}.
\newblock \showarticletitle{Augmented agile: Human-centered {AI}-assisted software management}.
\newblock \bibinfo{journal}{\emph{IEEE Software}} \bibinfo{volume}{40}, \bibinfo{number}{4} (\bibinfo{year}{2023}), \bibinfo{pages}{106--109}.
\newblock


\bibitem[Hoda et~al\mbox{.}(2018)]%
        {hoda2018rise}
\bibfield{author}{\bibinfo{person}{Rashina Hoda}, \bibinfo{person}{Norsaremah Salleh}, {and} \bibinfo{person}{John Grundy}.} \bibinfo{year}{2018}\natexlab{}.
\newblock \showarticletitle{The rise and evolution of agile software development}.
\newblock \bibinfo{journal}{\emph{IEEE Software}} \bibinfo{volume}{35}, \bibinfo{number}{5} (\bibinfo{year}{2018}), \bibinfo{pages}{58--63}.
\newblock


\bibitem[Hou et~al\mbox{.}(2024)]%
        {hou2024large}
\bibfield{author}{\bibinfo{person}{Xinyi Hou} {et~al\mbox{.}}} \bibinfo{year}{2024}\natexlab{}.
\newblock \showarticletitle{Large language models for software engineering: A systematic literature review}.
\newblock \bibinfo{journal}{\emph{ACM Transactions on Software Engineering and Methodology}} \bibinfo{volume}{33}, \bibinfo{number}{8} (\bibinfo{year}{2024}), \bibinfo{pages}{1--79}.
\newblock


\bibitem[Kruchten(2004)]%
        {kruchten2004rational}
\bibfield{author}{\bibinfo{person}{Philippe Kruchten}.} \bibinfo{year}{2004}\natexlab{}.
\newblock \bibinfo{booktitle}{\emph{The rational unified process: an introduction}}.
\newblock \bibinfo{publisher}{Addison-Wesley Professional}.
\newblock


\bibitem[Lago et~al\mbox{.}(2015)]%
        {lago2015framing}
\bibfield{author}{\bibinfo{person}{Patricia Lago} {et~al\mbox{.}}} \bibinfo{year}{2015}\natexlab{}.
\newblock \showarticletitle{Framing sustainability as a property of software quality}.
\newblock \bibinfo{journal}{\emph{Commun. ACM}} \bibinfo{volume}{58}, \bibinfo{number}{10} (\bibinfo{year}{2015}), \bibinfo{pages}{70--78}.
\newblock


\bibitem[Li et~al\mbox{.}(2025)]%
        {li2025rise}
\bibfield{author}{\bibinfo{person}{Hao Li}, \bibinfo{person}{Haoxiang Zhang}, {and} \bibinfo{person}{Ahmed~E Hassan}.} \bibinfo{year}{2025}\natexlab{}.
\newblock \showarticletitle{The Rise of {AI} Teammates in Software Engineering (SE) 3.0: How Autonomous Coding Agents Are Reshaping Software Engineering}.
\newblock \bibinfo{journal}{\emph{arXiv preprint arXiv:2507.15003}} (\bibinfo{year}{2025}).
\newblock


\bibitem[Lu et~al\mbox{.}(2022)]%
        {lu2022software}
\bibfield{author}{\bibinfo{person}{Qinghua Lu}, \bibinfo{person}{Liming Zhu}, \bibinfo{person}{Xiwei Xu}, \bibinfo{person}{Jon Whittle}, \bibinfo{person}{David Douglas}, {and} \bibinfo{person}{Conrad Sanderson}.} \bibinfo{year}{2022}\natexlab{}.
\newblock \showarticletitle{Software engineering for responsible {AI}: An empirical study and operationalised patterns}. In \bibinfo{booktitle}{\emph{Proceedings of the 44th International Conference on Software Engineering: Software Engineering in Practice}}. \bibinfo{pages}{241--242}.
\newblock


\bibitem[Mikalonyt{\.e} and Kneer(2022)]%
        {mikalonyte2022can}
\bibfield{author}{\bibinfo{person}{Elz{\.e}~Sigut{\.e} Mikalonyt{\.e}} {and} \bibinfo{person}{Markus Kneer}.} \bibinfo{year}{2022}\natexlab{}.
\newblock \showarticletitle{Can Artificial Intelligence make art?: Folk intuitions as to whether {AI}-driven robots can be viewed as artists and produce art}.
\newblock \bibinfo{journal}{\emph{ACM transactions on human-robot interaction}} \bibinfo{volume}{11}, \bibinfo{number}{4} (\bibinfo{year}{2022}), \bibinfo{pages}{1--19}.
\newblock


\bibitem[Perkusich et~al\mbox{.}(2020)]%
        {perkusich2020intelligent}
\bibfield{author}{\bibinfo{person}{Mirko Perkusich} {et~al\mbox{.}}} \bibinfo{year}{2020}\natexlab{}.
\newblock \showarticletitle{Intelligent software engineering in the context of agile software development: A systematic literature review}.
\newblock \bibinfo{journal}{\emph{Information and Software Technology}}  \bibinfo{volume}{119} (\bibinfo{year}{2020}), \bibinfo{pages}{106241}.
\newblock


\bibitem[Roychoudhury(2025)]%
        {roychoudhury2025thoughts}
\bibfield{author}{\bibinfo{person}{Abhik Roychoudhury}.} \bibinfo{year}{2025}\natexlab{}.
\newblock \showarticletitle{Agentic {AI} for Software: thoughts from Software Engineering community}.
\newblock \bibinfo{journal}{\emph{arXiv preprint arXiv:2508.17343}} (\bibinfo{year}{2025}).
\newblock


\bibitem[Roychoudhury et~al\mbox{.}(2025)]%
        {roychoudhury2025agentic}
\bibfield{author}{\bibinfo{person}{Abhik Roychoudhury}, \bibinfo{person}{Corina Pasareanu}, \bibinfo{person}{Michael Pradel}, {and} \bibinfo{person}{Baishakhi Ray}.} \bibinfo{year}{2025}\natexlab{}.
\newblock \showarticletitle{Agentic {AI} software engineer: Programming with trust}.
\newblock \bibinfo{journal}{\emph{Commun. ACM}} (\bibinfo{year}{2025}).
\newblock
\urldef\tempurl%
\url{arXiv preprint arXiv:2502.13767}
\showURL{%
\tempurl}


\bibitem[Russo et~al\mbox{.}(2024)]%
        {russo2024generative}
\bibfield{author}{\bibinfo{person}{Daniel Russo} {et~al\mbox{.}}} \bibinfo{year}{2024}\natexlab{}.
\newblock \showarticletitle{Generative {AI} in software engineering must be human-centered: The copenhagen manifesto}.
\newblock \bibinfo{journal}{\emph{Journal of Systems and Software}}  \bibinfo{volume}{216} (\bibinfo{year}{2024}).
\newblock


\bibitem[Sapkota et~al\mbox{.}(2025)]%
        {sapkota2025ai}
\bibfield{author}{\bibinfo{person}{Ranjan Sapkota}, \bibinfo{person}{Konstantinos~I Roumeliotis}, {and} \bibinfo{person}{Manoj Karkee}.} \bibinfo{year}{2025}\natexlab{}.
\newblock \showarticletitle{{AI} agents vs. agentic {AI}: A conceptual taxonomy, applications and challenges}.
\newblock \bibinfo{journal}{\emph{arXiv preprint arXiv:2505.10468}} (\bibinfo{year}{2025}).
\newblock


\bibitem[Seaman et~al\mbox{.}(2025)]%
        {seaman2025qualitative}
\bibfield{author}{\bibinfo{person}{Carolyn Seaman}, \bibinfo{person}{Rashina Hoda}, {and} \bibinfo{person}{Robert Feldt}.} \bibinfo{year}{2025}\natexlab{}.
\newblock \showarticletitle{Qualitative research methods in software engineering: past, present, and future}.
\newblock \bibinfo{journal}{\emph{IEEE Transactions on Software Engineering}} (\bibinfo{year}{2025}).
\newblock


\bibitem[Spiegler et~al\mbox{.}(2025)]%
        {spiegler2025images}
\bibfield{author}{\bibinfo{person}{Simone Spiegler}, \bibinfo{person}{Rashina Hoda}, {and} \bibinfo{person}{Aastha Pant}.} \bibinfo{year}{2025}\natexlab{}.
\newblock \showarticletitle{Images of {AI}: How {AI} practitioners view the impact of Artificial Intelligence on society, now and in the future}.
\newblock \bibinfo{journal}{\emph{Technology in Society}} (\bibinfo{year}{2025}).
\newblock


\bibitem[State of Agile report(2025)]%
        {stateOfAgile}
State of Agile report \bibinfo{year}{2025}\natexlab{}.
\newblock
\newblock
\shownote{\url{https://digital.ai/resource-center/analyst-reports/18th-state-of-agile-report/}}.


\bibitem[Storey et~al\mbox{.}(2025)]%
        {storey2025guiding}
\bibfield{author}{\bibinfo{person}{Margaret-Anne Storey}, \bibinfo{person}{Rashina Hoda}, \bibinfo{person}{Alessandra Maciel Paz~Milani}, {and} \bibinfo{person}{Maria~Teresa Baldassarre}.} \bibinfo{year}{2025}\natexlab{}.
\newblock \showarticletitle{Guiding principles for mixed methods research in software engineering}.
\newblock \bibinfo{journal}{\emph{Empirical Software Engineering}} \bibinfo{volume}{30}, \bibinfo{number}{5} (\bibinfo{year}{2025}), \bibinfo{pages}{138}.
\newblock


\bibitem[Wang et~al\mbox{.}(2025b)]%
        {wang2025ai}
\bibfield{author}{\bibinfo{person}{Huanting Wang}, \bibinfo{person}{Jingzhi Gong}, \bibinfo{person}{Huawei Zhang}, \bibinfo{person}{Jie Xu}, {and} \bibinfo{person}{Zheng Wang}.} \bibinfo{year}{2025}\natexlab{b}.
\newblock \showarticletitle{{AI} agentic programming: A survey of techniques, challenges, and opportunities}.
\newblock \bibinfo{journal}{\emph{arXiv preprint arXiv:2508.11126}} (\bibinfo{year}{2025}).
\newblock


\bibitem[Wang et~al\mbox{.}(2025a)]%
        {wang2025agents}
\bibfield{author}{\bibinfo{person}{Yanlin Wang} {et~al\mbox{.}}} \bibinfo{year}{2025}\natexlab{a}.
\newblock \showarticletitle{Agents in software engineering: Survey, landscape, and vision}.
\newblock \bibinfo{journal}{\emph{Automated Software Engineering}} \bibinfo{volume}{32}, \bibinfo{number}{2} (\bibinfo{year}{2025}), \bibinfo{pages}{1--36}.
\newblock


\bibitem[Xiao et~al\mbox{.}(2025)]%
        {xiao2025ai}
\bibfield{author}{\bibinfo{person}{Qing Xiao} {et~al\mbox{.}}} \bibinfo{year}{2025}\natexlab{}.
\newblock \showarticletitle{{AI} Hasn't Fixed Teamwork, But It Shifted Collaborative Culture: A Longitudinal Study in a Project-Based Software Development Organization (2023-2025)}.
\newblock \bibinfo{journal}{\emph{arXiv preprint arXiv:2509.10956}} (\bibinfo{year}{2025}).
\newblock


\bibitem[Zhang et~al\mbox{.}(2025)]%
        {zhang2025empowering}
\bibfield{author}{\bibinfo{person}{Sai Zhang} {et~al\mbox{.}}} \bibinfo{year}{2025}\natexlab{}.
\newblock \showarticletitle{Empowering agile-based generative software development through human-{AI} teamwork}.
\newblock \bibinfo{journal}{\emph{ACM Trans. on SE and Methodology}} (\bibinfo{year}{2025}).
\newblock


\end{thebibliography}

\end{document}